\documentclass[prl,twocolumn,longbibliography,letterpaper,showpacs,groupedaddress,superscriptaddress,nofootinbib,floatfix,preprintnumbers,tightenlines]{revtex4-1}
\usepackage[colorlinks,urlcolor=blue]{hyperref}
\usepackage{amsmath,amssymb,amsfonts}
\usepackage{graphicx}
\usepackage{bm}
\usepackage{comment}
\usepackage{color}
\usepackage{csquotes}
\usepackage{longtable}
\usepackage[section]{placeins}

\allowdisplaybreaks
\let\Oldsection\section
\renewcommand{\section}{\FloatBarrier\Oldsection}
\let\Oldsubsection\subsection
\renewcommand{\subsection}{\FloatBarrier\Oldsubsection}

\newcount\hour \newcount\hourminute \newcount\minute
\hour=\time \divide \hour by 60
\hourminute=\hour \multiply \hourminute by 60
\minute=\time \advance \minute by -\hourminute
\newcommand{\mydate}{\ \today \ - \number\hour :\number\minute}
\newcommand{\bea}{\begin{eqnarray}}
\newcommand{\eea}{\end{eqnarray}}

\newcommand{\eq}[1]{Eq.~(\ref{#1})}

\def\r{\rho}\def\a{\alpha}\def\d{\delta}\def\bfb{{\bf b}}\def\l{\lambda}\def\th{\theta}

\def\OMIT#1{{}}

\def\D{\Delta}
\def\bfr{{\bf r}}\def\bfb{{\bf b}}\def\bfk {{\bf k}}

\def\r{\rho}
\def\ra{\rangle}

\def\s{\sigma}

\begin{document}

\title{
The  Entanglement   of  Elastic and Inelastic Scattering}

\preprint{NT@UW-23-07}

\author{Gerald A.~Miller}
\affiliation{Department of Physics, University of Washington, Seattle, WA 98195-1560, USA}

\date{\mydate}

\begin{abstract}
  The entanglement properties of systems in which elastic and inelastic reactions occur in projectile-target interactions is studied. A new measure of entanglement, the scattering entropy, based on the unitarity of the $S-$matrix (probability conservation), is suggested. Using simple models for both low- and high-energy interactions,
the amount of entanglement is found to track with the strength of the inelastic interaction. The familiar example of the classical ``black disk", total absorption,  model is found to correspond to maximum entanglement. An analysis of high-energy $pp$ scattering data shows that entanglement is near maximum for lab energies greater than about 1 GeV, showing that the total absorption model is a reasonable starting point for understanding the data. 
   \end{abstract}
\maketitle

 \section{Introduction}


The implications of entanglement in quantum mechanics and quantum field theory have recently been studied in many papers.
For a long list of recent references see Ref.~\cite {Ehlers:2022oal}.
This  new interest  has been stimulated by the connection with quantum computing. Work related to hadron, QCD and  EIC  physics appears in
Refs.~\cite{Kharzeev:2017qzs,PhysRevC.99.015205,PhysRevD.102.074008,Hentschinski:2022rsa,PhysRevD.91.054026,Beane:2019loz}.  The entanglement properties of  nucleon-nucleon scattering and nucleon-nucleus elastic scattering are discussed in 
Refs.~\cite{Beane:2018oxh,Liu:2022grf,Miller:2023ujx,Bai:2023tey,Bai:2022hfv}. The connections between entanglement and  nuclear structure are presented in 
\cite{Johnson:2022mzk,Kruppa:2020rfa,Kruppa:2021yqs,Robin:2020aeh,Pazy:2022mmg,Tichai:2022bxr,Gu:2023aoc,Bulgac:2022ygo,Bulgac:2022cjg}.
There is also a  possible deep connection  between entanglement and underlying symmetries of the 
Standard Model~\cite{Cervera-Lierta:2017tdt,Beane:2018oxh,Liu:2022grf,Bai:2023tey,Miller:2023ujx}.
  
  The present paper is concerned with situations in which a projectile can  excite a target.   One of the challenges in studying entropy and entanglement  for scattering  is the need to develop proper definitions for the necessary infinite dimensional Hilbert space. This is done  here using the requirements of unitarity.

A special and somewhat ubiquitous case is the scattering of a  particle from a   totally absorbing ``black disk" of radius $R$~\cite{Kovchegov:2012mbw,Frauenfelder:1975dx,Gottfried2018}.   This situation approximately occurs in low-energy $\a$-nucleus scattering, and in high-energy proton-proton scattering.  In the  total absorption  limit, following the requirement of unitarity of the $S$-matrix,  the elastic $\s_{\rm el}$  and inelastic $\s_{\rm inel}$  cross sections are equal. The inelastic cross section cross section,  is $\pi R^2$, so that the total cross section is 2$\pi R^2$, twice the geometric cross section. I will argue that when $\s_{\rm el}=\s_{\rm inel}$ the entanglement entropy is maximized.

  \section{ Low-energy projectile-target scattering 
  and a new measure of entropy}
Consider projectile-target  scattering at energies sufficient low so that there is only $s$-wave scattering. Furthermore, the model definition is that there  is only  inelastic scattering to a single excited state, $X$. 
I  consider examples in which the inelastic scattering ranges from relatively small, corresponding, for example,  to neutron-nucleus scattering, 
to relatively large, corresponding to alpha-nucleus scattering. Another example, discussed below,  is nucleon-nucleon scattering in which interactions cause  either the target or projectile to be in an excited state.

The initial state is a product of a plane wave state and the target ground state, $G$. As a product state there is no entanglement. 
Interactions occur such that after the scattering event the projectile-target wave function is given by
\bea
 |\Psi\ra= |u_1\ra \otimes |G\ra+|u_2\ra \otimes |X\ra, \label{Sd1}
\eea
where $|u_1\ra$ represents a projectile with energy corresponding to elastic scattering and $|u_2\ra$ represents a projectile with an energy corresponding to inelastic scattering. Measurement of the energy of the projectile  determines whether or not the nucleus is in its ground or excited state. Thus the state represented by  \eq{Sd1} is an entangled state.  The next step is to work out a way to calculate entanglement properties.  The wave function, $|\Psi\ra$  is almost of  the form of the Schmidt decomposition in which the different coefficients represent probability amplitudes.  Here  the wave functions are in the continuum, so that discrete normalization conventions are not applicable.  It seems  necessary to develop a new method to compute entropy.

The procedure   is to use an  exactly soluble model ~\cite{Kamal:1970xy} to illustrate and develop the necessary formalism. I argue below  that the formalism is more general than the model. In this model the interactions are represented by delta-shell interactions~\cite{Gottfried2018} that can be thought of as approximating projectile-target  interactions at the surface of the target.  Then
the radial wave functions $u_{1,2}(r)$ satisfy the coupled-channels equations:
\bea
d^2u_1/ dr^2+[k^2-V_1\d(r-a)]u_1=V_{12}\d(r-a)u_2,\label{cc}\\
d^2u_2/ dr^2+[k^2-\D^2-V_2\d(r-a)]u_1=V_{21}\d(r-a)u_2.\label{cc1}
\eea
Hermiticity demands $V_{12}=V_{21}$ and  calculations are limited to the case $V_1\ne0,\,V_{12}\ne0,V_2=0$
to gain analytic insight. The parameter $\D$ is proportional to the energy difference between the excited and ground states.
The solution of \eq{cc} for $u_1$ is expressed in terms of the free-particle Green's function $g_1(r,r') $ as
\bea 
u_1(r)={\sin{k r}\over k} +V_1g_1(r,a)u_1(a) +V_{12}g_1(r,a)u_2(a),\nonumber\\
\label{u1}\eea
with 
\bea
g_1(r,r')=-(1/k)\sin{kr_<}e^{i k r_>},
\eea
 $r_<(r_>) $ is the smaller (larger) of  $(r,r') $.
The solution of \eq{cc1} for $u_2$ is given by
\bea
u_2(r)=V_{12}g_2(r,a)u_1(a)\label{u2}\eea
with 
\bea
g_2(r,r')=-(1/k_2)\sin{k_2r_<}e^{i k_2r_>},
\eea
where $k_2\equiv \sqrt{k^2-\D^2}$.
The results for $u_{1,2}(r)$ express the condition that the initial state is a plane wave incident on the ground state of the target nucleus.
The use of \eq{u2} in \eq{u1} leads to the result
\bea
u_1(r)=(1/k)\sin{k r}+ T_{11}e^{ik r}
\eea
for $r>a$, with the $T$-matrix element given by 
\bea
T_{11}={({\sin{ka}\over k})^2[V_1+V_{12}^2g_2(a,a)]\over1-[V_1+V_{12}^2g_2(a,a)]g_1(a,a)}.
\label{T11}\eea
The relation between $T_{11}$ and the complex-valued scattering phase shift, $\d_0$, is given by
\bea 
T_{11}={e^{2i\d_0}-1\over 2i k}.
\label{phase}\eea
Similarly
\bea u_2(r)=T_{12}e^{ik_2r},\eea
with 
\bea 
T_{12}={V_{12}({\sin{ka}\over k})({\sin{k_2a}\over k_2})\over1-[V_1+V_{12}^2g_2(a,a)]g_1(a,a)}.
\eea

Next,  turn to the entanglement properties of the model. The textbook definition 
is the entanglement entropy, the von Neumann entropy, given by
$S=-{\rm Tr}[\r \log_2\r],$
where $\r$ is the one-body density matrix.  This is typically evaluated by diagonalizing $\r$ in a discrete basis. Here  continuum wave functions,  normalized as delta functions, are used. So there is a need to obtain an appropriate definition of probability. This is done through the optical theorem, an expression of the unitarily of the $S$-matrix:
\bea
\s_{tot}={4\pi\over k_1} {\rm Im}[T_{11}].
\eea
The left-hand side is sum of the elastic and inelastic scattering cross sections, integrated  over all angles. The result  
for  the present model is expressed as
\bea
1={k_1 |T_{11}|^2 +k_2|T_{12}|^2\over {\rm Im}[T_{11}]}, \label{opt}
\eea
a relation that can be checked using the expressions for $T_{11}$ and $T_{12}$.
The \eq{opt} leads to a natural definition of probabilities based on the number of counts detected at an asymptoticaly located detector.
The ground state probability $P_G$ is given by
\bea P_G={k_1 |T_{11}|^2  \over {\rm Im}[T_{11}]}\label{PG}\eea and the excited state probability $P_X$ is given by
\bea
P_X={k_2 |T_{12}|^2  \over {\rm Im}[T_{11}]},\label{PX}\eea 
and via \eq{opt}: $P_G+P_X=1$.

Therefore, one may define the projectile-target  $(pT)$ entanglement entropy $S_{pT}$ of the final state as
\bea
S_{pT}=-P_G \ln_2P_G - P_X\ln_2 P_X.
\label{e1}
\eea
This entanglement entropy, termed the {\it scattering entropy}, is minimized if either of $P_G$ or $P_X$ vanishes. In that case the final scattering state is a simple tensor product. The  scattering entropy is maximized at $S_{pT}=1$ when $P_G=P_X.$  Note also that  \eq{phase} shows that $T_{11}$ is periodic in $k$, vanishing whenever $k=n\pi$.

\begin{figure}[h]
  \centering
 \includegraphics[width=0.4\textwidth]{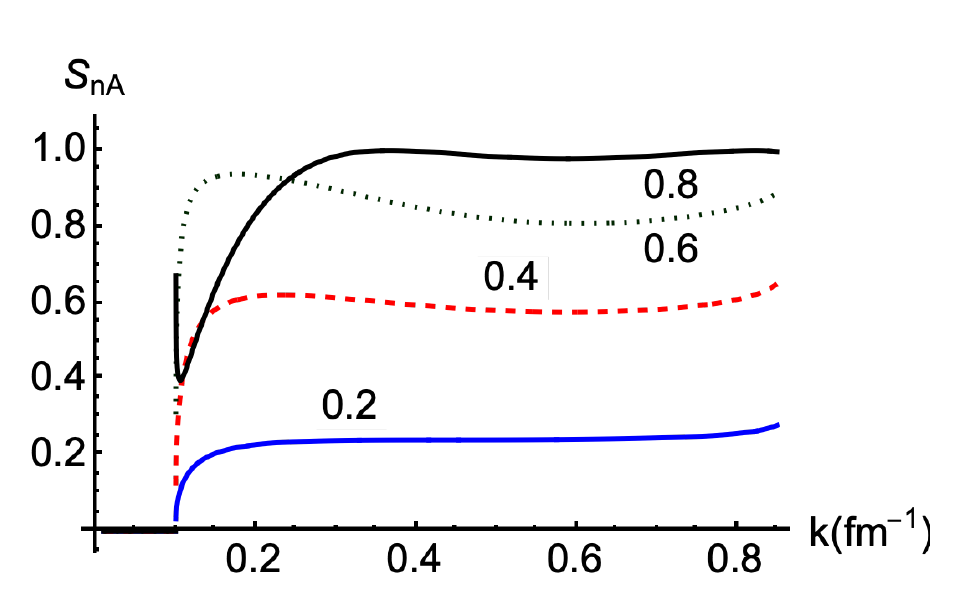}
  \hspace{0.3cm}
  \caption{  $S_{nA}$ as a function of $k=k_1$ for the four different values of $V_{12}/V$ shown in the figure.}
\label{SnA}  \end{figure}
Fig.~\ref{SnA} shows $S_{pT}$ for parameters $a=3.5 \,\rm fm , V_1=0.25 \,\rm fm^{-1}$ for different ratios $V_{12}/V_1$ as a function of $k$ the incident momentum.  The parameter $\D=0.1$ fm$^{-1}$.  The situation of $V_{12}/V_1=0.2$ is similar to that of neutron-nucleus interactions in which the inelastic scattering is relatively small.  The stronger absorption situation of $V_{12}/V_1=1$ is similar to that of alpha-nucleus interactions in which the inelastic scattering is large. 

For values of $k<\D$ the entanglement  entropy vanishes because the target cannot be excited. For higher values 
the scattering entropy is  at its maximum value when  $V_{12}/V_1=1$. This result  can be understood directly from \eq{PG} and \eq{PX}.  These quantities are approximately equal if $V/(ka)\ll1$ and $k\gg\D$. This result is similar to that of the total absorption  model in which the elastic and inelastic cross sections are the same. But here there is only one phase shift.  The unusual cusp-like near-threshold behavior for the case when $V_{12}/V_1=1$ arises  from the non-analytic square root behavior of $k_2$ combined with the increasing importance of the second term in the numerator of \eq{T11}.

The key  lesson of Fig.~\ref{SnA} is that entanglement entropy, as measured by the scattering entropy, increases as the tendency for inelastic scattering increases.



  \section{High energy scattering in a two-channel model}
 The scattering wave function $|\Psi\ra$ is given again by \eq{Sd1}.
In the high energy limit
the wave number $k$ is large compared to the inverse size of the system and large compared to the energy difference between the ground and excited states represented by $\D$. Thus
 $\D$ is neglected in solving the relevant wave equations, but kept  as very small, but non-zero, to  maintain the entanglement property that measuring energy of the projectile in the final state determines whether or not the target is in the ground state.  


The coupled-channel equations for high-energy scattering are then given by
\bea
\nabla^2\psi_1 +(k^2-V)\psi_1=U\psi_2\\
\nabla^2\psi_2 +(k^2-V)\psi_2=U\psi_1
\eea
The implementation of the eikonal or short-wavelength approximation is made by using $\psi_{1,2}(\bfr)=e^{i kz}\phi_{1,2}(\bfb,z)$ in which the direction of the beam is denoted as $\hat z$ and the direction transverse to that by $\bfb$. The procedure~\cite{Glauber}  is to use these in the coupled-channel equations and with large $k$ neglect the terms  $\nabla^2\phi_{1,2}$. This approximation is valid under two conditions~\cite{Glauber}: (i)  the short-wavelength limit that $1/k$ is less than any distance scale in the problem, and (ii) $(V,U)/k^2\ll1$ to prevent back-scattering. Then the coupled-channel equations become
\bea
2i k {\partial \phi_1 \over \partial z}-V\phi_1=U\phi_2\\
2i k {\partial \phi_2 \over \partial z}-V\phi_2=U\phi_1.
\eea
Let $\phi\equiv \phi_1+\phi_2$ and $\chi
\equiv \phi_1-\phi_2$. Adding the two equations gives
\bea 2i k {\partial \phi \over \partial z}=(U+V)\phi,\eea
and subtracting the two gives
\bea 2i k {\partial \chi \over \partial z}=(V-U)\phi\eea
with solutions
\bea\phi(\bfb,z)=\exp[{{-i\over 2k}\int_{-\infty}^z dz'(V(\bfb,z')+U(\bfb,z'))}\\
\chi(\bfb,z)=\exp[{{-i\over 2k}\int_{-\infty}^z dz' (V(\bfb,z')-U(\bfb,z'))}.
\eea
The two-component scattering amplitude is given by
\bea
\hat f(\bfk',\bfk)={-1\over 4\pi}\int d^3re^{-i\bfk'\cdot \bfb}\begin{bmatrix}
V&U\\
U&V
\end{bmatrix}\begin{bmatrix}\phi_1\\ \phi_2\end{bmatrix},
\eea
with the upper row of $\hat f$, $f_G$, corresponding to elastic scattering and the lower row, $f_X$, to inelastic scattering.
Then evaluation leads to the results
\bea
&
%
f_G(\bfk',\bfk)={i k\over 2\pi}\int d^2b e^{-i\bfk'\cdot\bfb}(1-e^{-i\d_V(\bfb)}\cos\d_U(\bfb))\label{fG}\\&
f_X(\bfk',\bfk)={-k\over 2\pi}\int d^2b e^{-i\bfk'\cdot\bfb}e^{-i\d_V(\bfb)}\sin\d_U(\bfb),
\eea
where
\bea \d_V\equiv {1\over 2k}\int_{-\infty}^\infty dz'V(\bfb,z'),\,\d_U\equiv {1\over 2k}\int_{-\infty}^\infty dz'U(\bfb,z').\label{delta}\eea

The evaluation of entanglement entropy requires an understanding  of 
unitarity. The statement of unitarity via the optical theorem is
\bea
\s_{Tot}=\int d\Omega(|f_G|^2+|f_X|^2)={4\pi\over k}\rm{Im}[f_G(\bfk',\bfk)],
\label{unit}\eea
a relationship that must be checked within the current model.
Taking the imaginary part of \eq{fG} yields
\bea
{\rm Im}[f_G(\bfk,\bfk)]={k\over 2\pi}\int d^2b (1-\cos{\d_V(\bfb)}\cos\d_U(\bfb)).
\eea
The evaluation of the 
 angular integrals of $|f_{G,X}|^2$ may be done using an approximation, valid when the eikonal approximation is valid, namely
\bea& \int d\Omega e^{i \bfk'\cdot(\bfb-\bfb')}\approx 2\pi {1\over k^2b }\d(b-b').\eea
Using this leads to the results
\bea &\int d\Omega |f_G(\bfk',\bfk)|^2=\int d^2b (1-2\cos\d_V\cos\d_U+\cos^2\d_U)\nonumber\\&
\int d\Omega |f_X(\bfk',\bfk)|^2=\int d^2b \sin^2\d_U,
\eea
so that
 the validity of \eq{unit} is maintained.
Therefore we may again define the  eikonal probability,  $P^e_{G,X}, $ as
\bea
&P_G^e={\int d^2b (1-2\cos\d_V(b)\cos\d_U(b)+\cos^2\d_U(b))\over 2 \int d^2b (1-\cos{\d_V(b)}\cos\d_U(b))}\\
&P_X^e={\int d^2b   \sin^2\d_U(b)\over  \int d^2b 2(1-\cos{\d_V(b)}\cos\d_U(b))},
\eea
and 
\bea
S^e=-P_G ^e\ln_2 P^e_G - P^e_X\ln_2 P^e_X.
\label{e2}
\eea

\begin{figure}[h]
  \centering
 \includegraphics[width=0.48\textwidth]{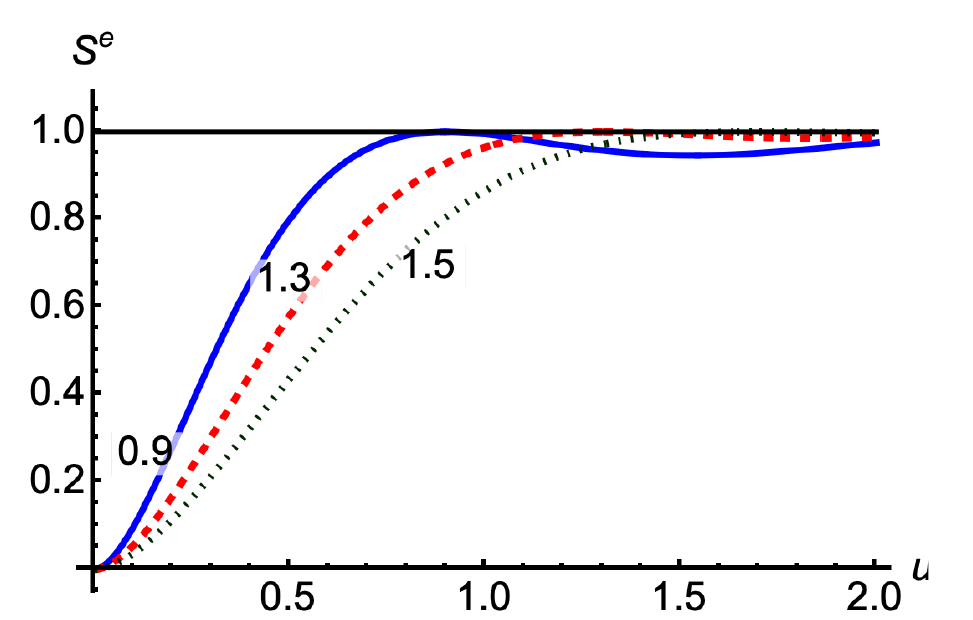}
  \caption{  $S^e$ as a function of  the dimensionless variable $u$ for the three  different values of $v$. The values of $v$ are 0.9 (solid), 1.1 (dashed) and 1.3 (dotted) . These values correspond to total cross sections of 22 mb, 40 mb and 56  mb.  
  }
\label{Glauber}  \end{figure}

The case with $U=\pm V$ yields $P_G^e=P_X^e=1/2$, and a maximum of entropy. This corresponds to the total absorption  limit in which elastic and inelastic cross sections are equal.
This means that the black disk limit corresponds to maximum scattering entropy.  

Presenting a brief   discussion of the total absorption limit is worthwhile. The partial wave decomposition of the scattering amplitude $f(\th)$ for a spinless particle is:
\bea f(\th)={-i\over 2k}\sum_l(2l+1)(\eta_l-1)P_l(\cos\th). 
\eea
The strong absorption model is defined by $\eta_l=0$ for $l\le L$ and $\eta_l=1$ for $l>L$, with $L\approxeq kR$. The sum is then
given by $f(\th)\approx {i\over k}L(L+1){J_1(L\th)\over L\th}$, a form familiar form Frauenhoffer diffraction.  In nuclear physics this is known as the Blair model~\cite{Blair:1954zz,TEO}.  See ~\cite{Abul-Magd:1969xiz}.  Data were reproduced using a distribution without a sharp edge, for example $\eta_l=1/(1+\exp{(L-l)/b})$ with $b>1/2$. This is a grey disc model.

To see if the total absorption or grey disc model is is a result of the present calculation,
I provide a specific example, based on parameters typical of proton-proton scattering Use a  Gaussian  density function $\r(r)=\exp((-r^2)/R^2)$, where $R$ is the radius parameter, taken here as $\sqrt{2}$ fm   obtained by convoluting Gaussian densities, of radius parameter 1 fm) of two protons.   Then  let $V(r)=V_0\r(r)$ and  $U(r)=U_0\r(r)$.  
Treating $u$ and $v$ as constants corresponds to treating the interactions as coming from vector exchanges-the typical treatment of high-energy hadron-hadron scattering.  The value of scattering entropy is then independent of energy for sufficiently high energies. In line with the high-energy behavior, I define 
$v\equiv 2\lambda_V k$ and $u\equiv 2\lambda_U k$ so that  evaluation of \eq{delta} yields the results  $\d_{V,U}(b)=\l_{V,U}\sqrt{\pi}R \exp(-b^2/R^2)$.
Then using \eq{unit}, a value of $\lambda_V$ of about 100 MeV gives a total cross section of about 40 mb, the typical value of the high-energy, proton-proton cross section. 

The results, independent of the signs of $U_0$ and $V_0$,  are shown in Fig.~2 in terms of  $u\equiv \l_{U}\sqrt{\pi}R $ and $v\equiv \l_{V}\sqrt{\pi}R $.
Maximum entanglement is reached, as expected, for cases with $u=v$.
Observe that, except for very small values of $u$ (small inelastic scattering) the entanglement entropy is always substantial.  

It is useful to learn if the results of the present calculation correspond to the total absorption or gray disc model.
To do this, refer to \eq{fG} and define $\eta(b)\equiv e^{-i\d_V(b)}\cos\d_U(b)$. This quantity is shown in Fig.~3 for the case $u=v=1.3$. The present calculation is seen to correspond to the grey disc model, not far from the total absorption model.

\begin{figure}[h]
  \centering
 \includegraphics[width=0.48\textwidth]{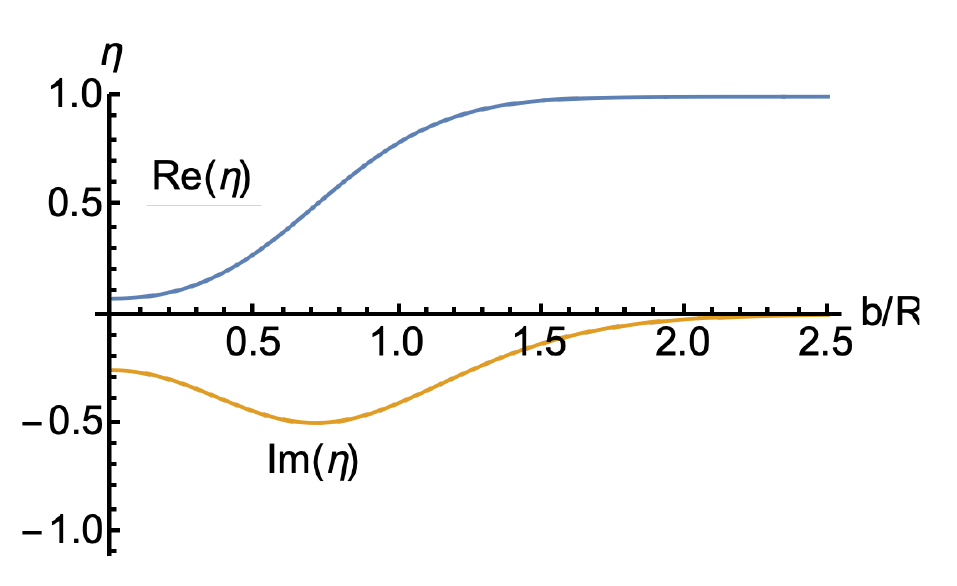} 
  \caption{  Real and imaginary parts of $\eta(b)$. 
 }
\end{figure}

 \section{Extension to more than one excited state and a general result}
 
Can  the models of the previous two models   be extended to include more than one excited state? What  then  can one say about entanglement?  If there is  more than one excited state,  a single measurement of the projectile energy cannot be used to determine the specific excited state of the target. The entanglement properties are then unknown. 
 
 However, a single measurement of the projectile energy can determine whether or not the target is excited.  Therefore it seems sensible to consider 
 the previous  terms $P_X$ and $P^e_X$ to  represent the probability that the target has been excited to any excited. In that case,
 the expressions for the scattering  entropy of \eq{e1} and \eq{e2} can be thought of as general measures of entanglement for any projectile-target system that involves inelastic excitation.
 
\section{high energy proton-proton scattering}
Data for total  cross sections and total elastic cross sections are available from the Particle Data Group~\cite{Workman:2022ynf}. Then,  the high-energy analysis presented above can be used with the identifications:
$P_G=\s_{\rm el}/\s_{\rm tot},\,P_X=1-P_G$ along with \eq{e2}.  The results are shown in Fig.~4.

\begin{figure}[h]
  \centering
 \includegraphics[width=0.48\textwidth]{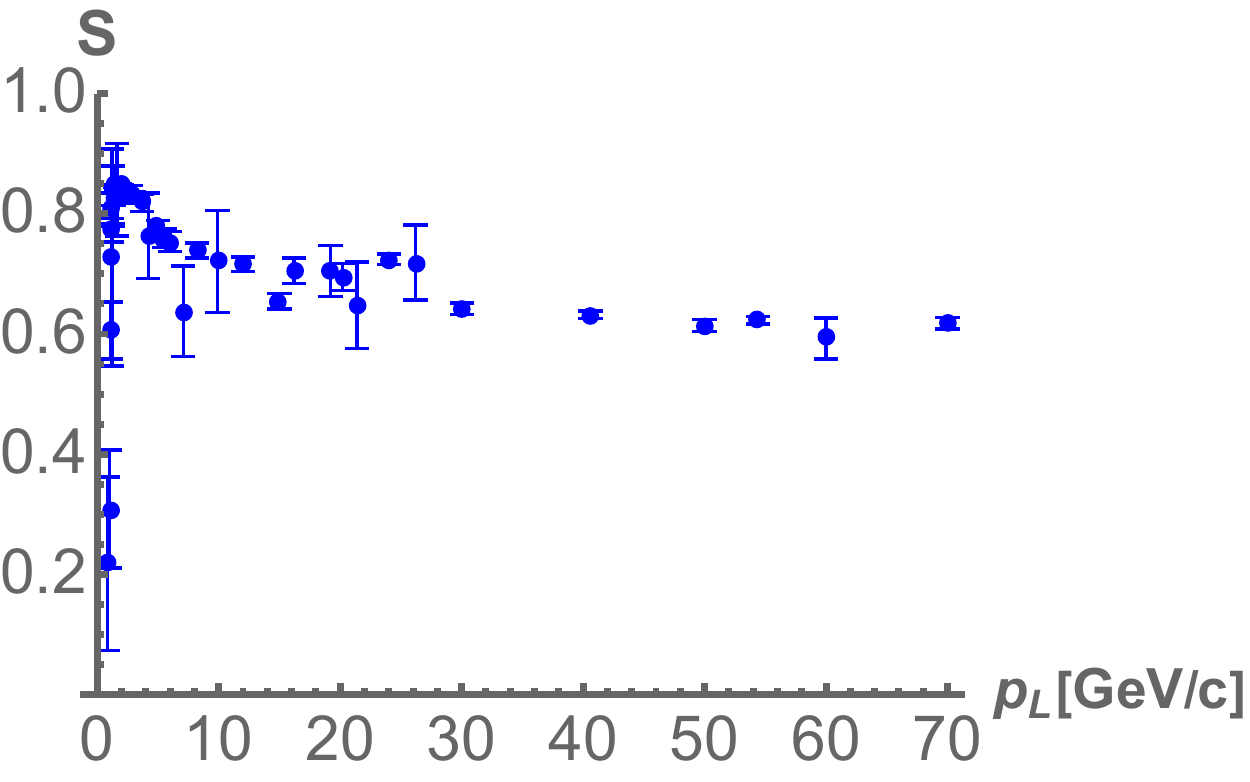} 
  \caption{  $S^e$ as a function of  the  fixed-target laboratory momentum $p_L$.. 
  }
\end{figure}

At low energies there is no inelastic scattering, so the scattering  entropy must vanish. This result is  similar to the results shown in Fig.~1 for small values of $k$. and to Fig.~2 for small values of $u$. As  energies rise above inelastic scattering thresholds the entanglement increases.  At still higher energies the ratio of elastic to total cross sections is approximately flat.  The entanglement entropy is substantial at laboratory momenta greater than about 2 GeV/c (kinetic energy about 1.3  GeV). At higher energies than are shown $S$ is approximately flat with energy because the ratio $\s_{|rm el}/\s_{\rm tot}$ is approximately independent of energy. 

The large value of entanglement entropy indicates that the total absorption or gray disc model  are reasonable first approximations to understanding the data. The net result is that  computing the entanglement energy provides insight regarding the underlying dynamics of proton-proton scattering, in particular and more generally of  projectile-target scattering.

\vskip.3cm
 This work was supported by the U. S. Department of Energy Office of Science, Office of Nuclear Physics under Award Number DE-FG02-97ER-41014. 
%

 \end{document}